# Empirical Fourier Decomposition

Wei Zhou [a]

Zhongren Feng [a, b]

Xiongjiang Wang [a, *]

Hao Lv [a]

[a] School of Civil Engineering and Architecture, Wuhan University of Technology, Wuhan 430070, P.R. China.

[b] School of Urban Construction, Wuchang Shouyi University, Wuhan 430064, P.R. China.

* Corresponding author. 122 Luoshi Road, Wuhan, 430070, P.R. China.

E-mail: wangxiongjiang@whut.edu.cn

**Abstract**

In this paper, a novel decomposition method for non-stationary and nonlinear signals is proposed. This method is inspired by the adaptive wavelet filter bank of the empirical wavelet transform (EWT) and Fourier intrinsic band functions (FIBFs) of the Fourier decomposition method (FDM). Therefore, the proposed approach is entitled as empirical Fourier decomposition (EFD). EFD is defined as the adaptive bandpass filter bank, regarded as the adaptive FIBFs based on the segment of the Fourier spectrum. Firstly, an enhanced segmentation technology of the Fourier spectrum based is presented. Secondly, the framework of EFD is established both in a continuous series and a discrete series. Finally, combined with the Hilbert transform, EFD is extended to a time-frequency representation. To verify the effectiveness of EFD, three non-stationary multimode signals, a simulated free vibration, and one real ECG signal are tested. The results manifest that EFD is more effective, compared with EWT and FDM, with higher processing precision, computation efficiency and noise robustness particularly to the closely-spaced frequencies and high-frequency noise.

# 1. Introduction

Multicomponent non-stationary and nonlinear (NSNL) signals decomposition has drawn deep concern in numerous fields, such as biomedical signal analysis [1–6], seismic signal analysis [7–10], vibration analysis in engineering [11–17], speech enhancement [18–21], etc. The multicomponent signals created by real-life physical systems frequently comprise several superposed oscillations, which are also termed as signal modes [22], and encompass meaningful information of a physical system. Hence, it is crucial to improve the accuracy and efficiency in signal processing.

During the past decades, several signal decomposition approaches have been proposed, among which empirical mode decomposition (EMD) has had a significant impact [23]. Despite the limited mathematical understanding and some obvious shortcomings, such as mode mixing and end effects, EMD is widely used in a broad variety of time-frequency analysis applications. To overcome these difficulties, some elevated versions of the EMD were developed, such as ensemble EMD [24], complete ensemble EMD [25], and recent method: time-varying filter based EMD [26]. Although these approaches abate the limitations to some extent, the performance is still not satisfactory as desired.

Jumping out of the bondage of EMD, several EMD-like signal processing methods have been proposed. One category of the method is retrieving each signal mode in time-frequency (TF) domain. These TF analysis methods represent the frequency of each mode versus time meaning that the time-varying frequency of non-stationary signals can be better dealt with [22]. Introduced by Daubechies [27], Synchrosqueezing transform (SST) is an EMD-like approach with combining wavelet analysis and reallocation technique. The reallocation method used in the SST enhances the resolution of the TF domain so tremendous that the SST is applied extensively. Subsequently, researchers proposed several SST-based methods, for instance, the higher order SST [28], the matching SST for signals with fast varying instantaneous frequency [17], time-reassigned SST with reassigned in the time direction [13], and local maximum SST for the energy concentrated in the TF domain greatly [12], and multi-SST via multiple reallocate to achieve high concentrated TF representation [29].

Another classify method is various adaptive filtering techniques. The empirical wavelet transform (EWT) employs the wavelet filter bank based on the segment of the Fourier spectrum [30]. To eliminate the restriction of the Fourier spectrum, the power spectrum based EWT methods have been developed [15,16,31]. The variational mode decomposition (VMD) belongs to this category and is a generalization of the Wiener filter [32]. Then, the method in [33] uses the adaptive local iterative filtering and the swarm decomposition algorithm which is seen as the swarm filtering [34]. In addition, the Fourier decomposition method (FDM) [35] and the recent adaptive chirp mode pursuit (ACMP) [14] are also the filter-based methods, but the FDM is based on the Fourier transform (FT) which is a TF filter bank with the adjustable bandwidth.

Some other methods utilize an optimization model to decompose signals. For instance, the operator-based approaches [36,37], the sparse TF method [38] and the short-time narrow-banded mode decomposition (STNBMD) [39]. Of course, the classification is not really rigid. For example, the frameworks of VMD and ACMP also contain optimization. However, the results of the optimization algorithms typically rely on the initialization, like the STNBMD, of which the initial frequencies influence the results deeply to the closely-spaced modals. Additionally, the bandwidth is regularly as the part of the objective function of the optimization methods, always predetermined [32] and narrow [39] in the early methods but recently the ACMP enables it adjustable.

In the aforementioned methods, two of them worth further researching, namely the EWT and the FDM. It is because both of them depend on the FT which has been regarded as inappropriate to analyze the NSNL signals for past decades [35]. Although the EWT method is the wavelet filter bank which is framed by the wavelet transform, the key of the EWT is the segment based on the Fourier spectrum [30]. Therefore, the EWT has great significance to spread the FT into the NSNL data analysis. Then, the FDM introduces the concept of Fourier intrinsic band functions (FIBFs) to decompose signals by employing the properties of the mode: non-negative amplitude and non-decreasing phase. The FDM applies the FT to the signals of NSNL authentically, based on the theoretic of the FT purely. Admittedly, EWT and FDM, especially EWT, have achieved great success and widely used in signal processing. These involved signal decomposition in the bearing fault diagnosis [11], structural modal analysis [15,16], medical analysis [1], to name just a few examples. Nonetheless, several limitations still persist in the EWT and FDM. With regard to the EWT, the transition phase of the wavelet filter bank may be redundant. It causes that the adjacent components interfere with each other, particularly the closely-spaced frequencies. For another, although it has been improved in subsequent literature, the method of the segmentation can continue to be optimized [40]. The current method cannot match the number of segments and the number of modes. Sometimes the first component is the residue and sometimes is the main component, which obscures the operator. FDM is sensitive to the noise. The noise signal can be decomposed too many components, moreover, the cost of the computation will increase drastically. Furthermore, the signal analyzed in FDM is not expanded, resulting in a very serious end effect in the decomposition components. These aforementioned defects all leave room for theoretical development and improvement on the robustness of the decomposition.

In this work, a novel signal decomposition algorithm called empirical Fourier decomposition (EFD) is proposed to address the hindrances aforementioned. Enlightened by EWT and FDM, EFD combines the idea of the segmentation in the Fourier spectrum and the FIBFs to decompose the signal into several modes. We propose a new method of partitioning and construct a mathematical framework of the EFD. The algorithm no longer possesses the complex transition phase. We also illustrate that the segmentation technique matches the number of segments and modes. Moreover, EFD optimizes performance under closely-spaced modes and high-frequency noise, and increases computation efficiency.

The remnant of the paper is arranged as follows. In section 2, we recall the principle of the three methods: EMD, EWT, and FDM. In section 3, we introduce the proposed method and expound its segmentation, frame of the continuous series, discrete series, and TF representation. Section 4 shows five experiments based on simulated and real signals. Finally, we present a conclusion for our work and some prospects for future works in section 5.

## 2. Theoretical background

*2.1 Empirical mode decomposition*

In 1998, EMD is proposed by Huang et al. [23], it is an adaptive algorithm to decompose a signal into several components. The components decomposed by this method are called intrinsic mode functions (IMFs). These IMFs are amplitude-modulated-frequency-modulated (AM-FM) signals, and so the original signal is written as:

$$f(t) = \sum_{k=1}^{N} f_k(t)\cos(\phi_k(t)) + r(t) \qquad (1)$$

where the envelop $f_k(t)$ is non-negative, the phase $\phi_k(t)$ is an increasing function, $r(t)$ is the final residue, and, $N$ is the number of IMF. $f_k(t)$ and instantaneous frequency $\omega_k(t) = \phi'_k(t)$ change markedly slower than $\phi_k(t)$ does [27,30,32].

All the IMFs should satisfy two fundamental conditions [23]: (a) The difference between the number of extreme points (minima and maxima) and the number of zero-crossings is no more than one over the entire duration of time series. (b) At any point, the mean of the envelope defined by the local minima and local maxima, respectively, is zero.

With these two basic conditions, the operational process of EMD is described as the following 5 steps:

1. Compute the upper and lower envelopes by using a cubic spline interpolation to fit the maxima and minima of $f(t)$.
2. Obtain the mean envelope and note $m(t)$.
3. Get a candidate IMF via $r_1(t) = f(t) - m(t)$. Generally, $r_1(t)$ dissatisfies the two above-mentioned conditions, then, repeat previous two steps to $r_k(t)$ ($k = 1, 2, \ldots, n-1$) until the eligible candidate $r_n(t)$ is found.
4. Define $r_n(t)$ as $f_1(t)$, and get the next IMF by employing the same steps to $(f(t)-f_1(t))$.
5. Repeat the aforementioned steps to calculate the remaining IMFs.

*2.2 Empirical wavelet transform*

EWT is an adaptive wavelet transform algorithm based on the segmentation of the Fourier spectrum [30]. This algorithm assumes that the Fourier support $[0, \pi]$ is divided into $N$ contiguous partitions. Each segment is denoted $S_n = [\omega_{n-1}, \omega_n]$ (the $\omega_n$ represents boundary, and $\omega_0 = 0$, $\omega_N = \pi$), then a transition zone $T_n$ is defined around $\omega_n$ symmetrically with the double width of $\tau_n$ (Fig. 1). The empirical wavelets are seen as one lowpass $\phi_n(\omega)$ and $N-1$ bandpass filters $\psi_n(\omega)$ on each $S_n$ [31].

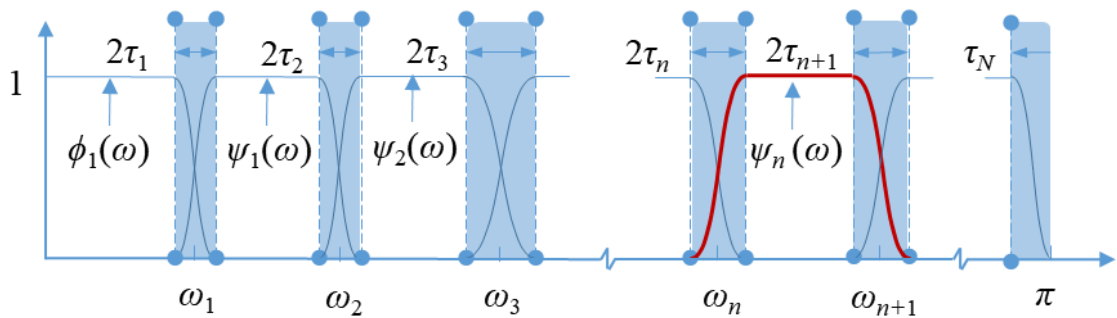

**Fig. 1. Basic construction of EWT.**

Fig. 1 exhibits the lowpass and bandpass wavelet filters defined as $\phi_1(\omega)$ and $\psi_n(\omega)$, where the horizontal axis and vertical axis correspond with the boundaries and amplitude of the filters, respectively. The first segment is the lowpass filter $\phi_1(\omega)$ with a cut-off frequency $\omega_1$, which corresponds to the first boundary.

Except for the last bandpass filter $\psi_N(\omega)$ with a frequency band $[\omega_{N-1}, \pi]$, other bandpass filters $\psi_n(\omega)$ have a frequency band of $[\omega_n, \omega_{n+1}]$.

The original segmentation technique in EWT is a local maxima approach [40]. In this method, Gilles computes all local maxima $M_i$ of the analyzed signal in the Fourier domain and deduces their corresponding position $\omega^i$. Then the $N$-1 largest maxima are set to all corresponding $\omega^i$ and re-index them as $\omega^n$ where $1 \leq n \leq N$-1. So, the set of boundaries $\mho = \{\omega_n\}_{n=0,\ldots,N}$ by:

$$\omega_n = \begin{cases} 0 & \text{if } n=0 \\ (\omega^n + \omega^{n-1})/2 & \text{otherwise} \\ \pi & \text{if } n=N \end{cases} \tag{2}$$

Next, following the idea utilized in originating both of Meyer's wavelets and Littlewood-Paley, the empirical scaling function and empirical wavelet function defined as Eqs. (3) and (4), respectively [41].

$$\phi_n(\omega) = \begin{cases} 1 & \text{if } |\omega| \leq \omega_n - \tau_n \\ \cos\left[\frac{\pi}{2}\beta\left(\frac{1}{2\tau_n}(\tau_n + |\omega| - \omega_n)\right)\right] \\ & \text{if } \omega_n - \tau_n \leq |\omega| \leq \tau_n + \omega_n \\ 0 & \text{otherwise} \end{cases} \tag{3}$$

and

$$\psi_n(\omega) = \begin{cases} 1 & \text{if } \omega_n + \tau_n \leq |\omega| \leq \omega_{n+1} - \tau_{n+1} \\ \cos\left[\frac{\pi}{2}\beta\left(\frac{1}{2\tau_{n+1}}(\tau_{n+1} + |\omega| - \omega_{n+1})\right)\right] \\ & \text{if } \omega_{n+1} - \tau_{n+1} \leq |\omega| \leq \tau_{n+1} + \omega_{n+1} \\ \sin\left[\frac{\pi}{2}\beta\left(\frac{1}{2\tau_n}(\tau_n + |\omega| - \omega_n)\right)\right] \\ & \text{if } \omega_n - \tau_n \leq |\omega| \leq \tau_n + \omega_n \\ 0 & \text{otherwise} \end{cases} \tag{4}$$

where $\tau_n = \gamma\omega$, read this paper [30] for more details about the value of $\gamma$. $\beta(x)$ is an arbitrary function with the following properties:

$$\beta(x) = \begin{cases} 0 & \text{if } x \leq 0 \\ & \text{and } \beta(x) + \beta(1-x) = 1 \quad x \in [0, 1] \\ 1 & \text{if } x \geq 1 \end{cases} \tag{5}$$

Moreover, in EWT, $\beta(x)$ is the same as the most utilized in [41]:

$$\beta(x) = x^4(35 - 85x + 70x^2 - 20x^3) \tag{6}$$

After the tight frame wavelet filters have been established, EWT can be defined by detail coefficients and approximation coefficients, respectively, as follows:

$$W_f^\varepsilon(n,t) = F^{-1}(f, \psi_n) \tag{7}$$

$$W_f^\varepsilon(0,t) = F^{-1}(f, \phi_1) \tag{8}$$

where $W_f^\varepsilon(n,t)$ is the detail coefficients, $W_f^\varepsilon(0,t)$ is the approximation coefficients, and $F^{-1}$ symbolizes the inverse Fourier transform.

### 2.3 Fourier decomposition method

Proposed by Singh [35], FDM is a novel adaptive approach to process the NSNL signal. This algorithm supposes a signal has $M$ FIBFs and uses the FIBF to build the mono-component of a signal. $M$ generally equals to the number of modes of a signal. The FIBF is identified as a real part of analytic Fourier intrinsic band functions (AFIBFs). The AFIBF is similar to a generalized Fourier expansion [23], but the limits of summation may vary for each AFIBF. Moreover, AFIBF is defined as:

$$\text{AFIBF}_i = a_i(t)\exp(j\varphi_i(t)) = \sum_{k=N_{i-1}+1}^{N_i} c_k \exp\left(\frac{2\pi jkt}{T_0}\right) \tag{9}$$

where $i = 1, 2, \ldots, M$ with $N_0 = 0$, $N_M = (N/2 - 1)$ when $N$ is even (or if $N$ is odd with $N_M = (N-1)/2$). $N_i$ is a vital value for AFIBF, essentially the boundary in the Fourier domain of FDM. It obtains adaptively by two strategies: low to high-frequency scan and high to low-frequency scan.

Subsequently, each method can be divided into four steps. Owing to these steps are similar except scan order, only the process of the low to high-frequency scan is described in the following paragraphs. For a complete comprehensible interpretation of the algorithm, please refer to the literature [35].

Step A. Obtain FFT of $f(t)$, i.e. $F(t) = \text{FFT}\{f(t)\}$.

Step B. Set $\text{AFIBF}_i = a_i(t)\exp(j\varphi_i(t)) = \sum_{k=N_{i-1}+1}^{N_i} c_k \exp\left(\frac{2\pi jkt}{T_0}\right)$.

Step C. Get maximum number of $N_i$ and a monotonically increasing function of phase $\varphi_i(n)$ which such that $(N_{i-1} + 1) \leq N_i \leq (N/2 - 1)$ and $\omega_i(n) = (\varphi_i(n+1) - \varphi_i(n-1))/2 \geq 0$, $\forall n$, respectively. Substitution $N_i$ and $\varphi_i(n)$ to step B to get the AFIBFs.

Step D. Obtain FIBFs from the real part of AFIBFs.

## 3. Proposed method

The proposed method, EFD, is can be considered as a partial combination of EWT and FDM. In the frame of this method, the ideas of segmenting Fourier spectrum in the EWT [30] and FIBF in FDM [35] are combined. EFD employs the segmentation of the Fourier spectrum, and each partition denotes the region of a FIBF. From Fig. 2, this method seems as a bandpass filter in each segment. Compared with EWT, EFD is more concise and not require the cumbersome transition phase and tight frame. Then, the following subsections describe EFD in four portions.

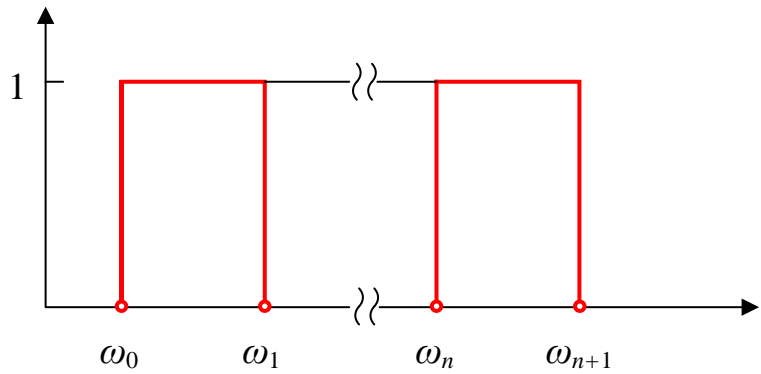

**Fig. 2. Basic construction of EFD.**

*3.1 Segmentation of the empirical Fourier decomposition*

As same to EWT, the segment of the Fourier spectrum is significant to EFD. High quality of segmentation is the prerequisite for the program of EFD. We propose a new segment method inspired by the local maxima method [30] and the local minima method [40].

For the purpose of the desired performance, the separation of the Fourier spectrum should be supported as tightly as possible. In this paper, the frequency range of the Fourier spectrum is also from 0 to $\pi$. The number of partitions is $N$, which means that a total of $N+1$ boundaries are necessitated. To determine these boundaries, the control points (the local maxima and initial value) are detected in the spectrum and sorted in descending order. Then, the assumption of identified $M$ control points is set as same to EWT [30]:

1. If $M \geq N$: the method found adequate control points to support the $N$ segments, and only the first $N$-1 control points are kept.
2. If $M < N$: the analyzed signal has fewer modes to extract, then keeping all the control points and

resetting $N$ to $M$.

Then, following the detected control point, their corresponding position $\omega^i$ is deduced. Next, re-index them as $\omega^n$ where $1 \leq n \leq N-1$, $\omega^0 = 0$ and $\omega^N = \pi$. Finally, we set the global minimum in $[\omega^{n-1}, \omega^n]$ as boundaries $\omega_n$, and the set of these minima is denoted as $\Omega_n$. So the Fourier boundaries is described as:

$$\omega_n = \begin{cases} \arg\min_{\omega} \Omega_n & \text{for } 1 \leq n \leq N-1 \\ \dfrac{\omega^{N-1} + \omega^N}{2} & n \leq N \end{cases} \quad (10)$$

Fig. 3 shows an example of the segmentation technique based on these three distinct methods. The example (solid line) the Fourier spectrum demonstrates that the signal has the following features. a) The first component of the signal concentrates in low frequency. b) The 2nd component is a flat-picked mode where the bandwidth is wide. c) The high-frequency noise exists in the signal. Figs. 3 (a), 3 (b) and 3 (c) show results of the local maxima method [30], the local minima method [40], and EFD. These results illustrate that the proposed method performs better for the signals with these three characteristics. Firstly, when the signal concentrates in the low frequency, the prior number of modes does not correspond accurately to the real number of modes in the local maxima method [30] and the local minima method [40]. In this case, the prior number of modes of the two methods is one less than the real number of modes. Secondly, for the flat-picked mode, the original EWT cannot decompose the modes well, which the 1st and 3rd components are contaminated with the 2nd. Thirdly, the proposed bound method alleviates the interference of high-frequency noise.

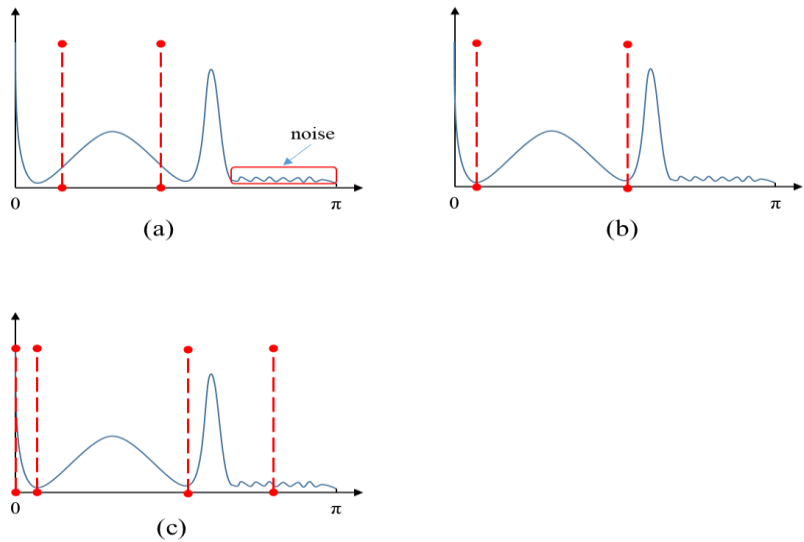

Fig. 3. Three segmentation methods: (a) the local maxima method, (b) the local minima method, (c) the proposed approach.

## 3.2 Continuous empirical Fourier decomposition

In this subsection, we construct a framework of continuous EFD. Firstly, let $f(t)$ be a signal of real-valued and time-limited ($t \in [t_1, t_1+L]$, $L>0$), and $\varphi_0:=2\pi/L$. So, the Fourier series expansion of $f(t)$ is determined as:

$$\begin{cases} a_0 = \frac{2}{L}\int_{t_1}^{t_1+L} f(t)\,dt \\ a_n = \frac{2}{L}\int_{t_1}^{t_1+L} \cos(n\varphi_0 t) f(t)\,dt \\ b_n = \frac{2}{L}\int_{t_1}^{t_1+L} \sin(n\varphi_0 t) f(t)\,dt \\ f(t) = \frac{a_0}{2} + \sum_{n=1}^{\infty}\left[a_n \cos(n\varphi_0 t) + b_n \sin(n\varphi_0 t)\right] \end{cases} \quad (11)$$

Then, using Euler's formula, the complex exponential representation of the Fourier series is given by:

$$f(t) = \frac{a_0}{2} + \frac{1}{2}\sum_{n=1}^{\infty} c_n \exp(jn\varphi_0 t) + c_n^* \exp(-jn\varphi_0 t) \quad (12)$$

where $c_n=(a_n-jb_n)$ and $c_n^*=(a_n+jb_n)$.

To give the definition of AFIBF in this proposed method, $u(t)$, its complex conjugate $\overline{u}(t)$ are set as:

$$\begin{cases} u(t) = \sum_{n=1}^{\infty} c_n \exp(jn\varphi_0 t) \\ \overline{u}(t) = \sum_{n=1}^{\infty} c_n^* \exp(-jn\varphi_0 t) \end{cases} \quad (13)$$

So Eq. (12) can be rewritten as:

$$f(t) = \frac{a_0}{2} + \text{Re}\{u(t)\} \quad (14)$$

where $\text{Re}\{u(t)\}$ is the real part of $u(t)$.

Following the segmentation in the subsection of 3.1, the number of segment ($N-1$) means that $u(t)$ can be separated into $N-1$ parts, and $u(t)$ can be computed as follows in a general Fourier expansion [23]:

$$u(t) = \sum_{i=1}^{N-1} A_i \exp(j\theta_i(t)) \quad (15)$$

Finally, combining the boundaries $\omega_n$, we describe AFIBFs as:

$$A_i \exp(j\theta_i(t)) = \sum_{\omega_i}^{\omega_{i+1}} c_n \exp(jn\varphi_0 t) \quad (16)$$

where $A_i$ is the instantaneous amplitude and $\theta_i$ is the instantaneous phase of the $i$-th FIBF, and FIBF is a real part of AFIBF.

### 3.3 Discrete empirical Fourier decomposition

In digital signal processing, discrete time series signals are regularly considered. Therefore, we present the framework of EFD for a discrete signal in this subsection. We set $x[n]$ as a discrete time series signal of length $K$, and employ the discrete FT to $x[n]$:

$$x[n] = \sum_{k=0}^{K-1} X[k] \exp\left(\frac{j2\pi kn}{K}\right) \tag{17}$$

where $X[k]=(1/K)\sum_{k=0}^{K-1} x[n]\exp(-j2\pi kn/K)$, which is the discrete FT of the signal $x[n]$. To segregate the Eq. (17) in symmetrically, $K$ is set to an even number, Then $X[0]$ and $X[K/2]$ are real numbers, and $x[n]$ is redefined as:

$$\begin{cases} x[n] = X[0] + X\left[\dfrac{K}{2}\right](-1)^n + 2\operatorname{Re}\{v[n]\} \\ v[n] = \sum_{k=1}^{K/2-1} X[k]\exp\left(\dfrac{j2\pi kn}{K}\right) \\ v^*[n] = \sum_{k=K/2+1}^{K-1} X[k]\exp\left(\dfrac{j2\pi kn}{K}\right) \end{cases} \tag{18}$$

where $\operatorname{Re}\{v[n]\}$ denote the real part of $v[n]$, and since $x[n]$ is real, $v[n]$ and $v^*[n]$ are complex conjugate.

Same to subsection 3.2, $v[n]$ in a general Fourier expansion is defined as:

$$v[n] = \sum_{k=1}^{K/2-1} X[k]\exp\left(\frac{j2\pi kn}{K}\right) = \sum_{i=1}^{N-1} A_i \exp(j\theta_i[t]) \tag{19}$$

Then, we give the AFIBFs in the discrete domain as:

$$A_i \exp(j\theta_i[t]) = \sum_{\omega_i}^{\omega_{i+1}} X[k]\exp\left(\frac{j2\pi kn}{K}\right) \tag{20}$$

### 3.4 Hilbert transform-empirical Fourier decomposition

For the analyzed signal presents more information, TF representation is contemplated. Using the idea employed in the Hilbert-Huang transform [23], we apply the Hilbert transform to the IMFs decomposed by EFD. Firstly, the definition of the Hilbert transform of a function $f$ is recalled as:

$$H_f(t) = \frac{1}{\pi} p.v. \int_{-\infty}^{+\infty} \frac{f(\tau)}{t-\tau} d\tau \tag{21}$$

where the integral calculus is defined by the value (p.v.) of the Cauchy principal [42].

By Hilbert transform, the analytical form $z$ of $f$ can be derived: $z(t)=f(t)+jH_f(t)$. With regard to AM-FM signals $f(t)$, Hilbert transform permits to give $z(t)=A(t)\exp(j\theta(t))$. In this method, the instantaneous amplitude $A(t)$ and frequency $\theta'(t)$ of the IMFs can be extracted which are decomposed through EFD. Moreover, in TF representation, each curve $\theta_i'(t)$ is plotted in the TF plane where the plot intensity is provided by $A_i(t)$.

Following this method, TF representation based on EFD is quite feasible.

## 4. Numerical studies and application

In this section, we use five examples: four artificial signals and one real-life signal, to test this proposed method. The form of the first three signals is utilized in the EWT [30] and VMD [32], and the fourth signal has been also applied in previous work [31]. Meanwhile, we chose EWT and FDM with applying to these signals for comparison. In these examples, the segments of EFD are set 4, 5, 3, 4 and 11 from the first signal to the fifth, respectively. The original EWT is used, and its selection of the segment is one less than EFD, except for the fourth example which is the same as EFD. Additionally, due to the similar results computing by these two methods in FDM, only the approach of low to high frequency is used in this paper.

*4.1 Non-stationary multimode signals*

Three non-stationary multimode signals, which primitively derive from the literature [38], are simulated in this section. All signals are in the region of [0, 1], and sample frequency is presupposed as 1000 Hz in this paper.

a) Example 1: The first signal has three simple components: one general linear trend and two diverse harmonics, their functions are described as follows:

$$\begin{cases} f_{11}(t) = 6t \\ f_{12}(t) = 2\cos(8\pi t) \\ f_{13}(t) = \cos(40\pi t) \\ f_1(t) = f_{11}(t) + f_{12}(t) + f_{13}(t) \end{cases} \tag{22}$$

The time series of $f_1(t)$ and its three components are shown in Fig. 4. For a direct exhibition of the segments of FDM, the boundaries are displayed in the same way as EFD and EWT, and all their boundaries are depicted in Fig. 5. It is worth noting that EFD is capable of separating the Fourier spectrum appropriately in $f_1(t)$. Owing to 0 and $\pi$ are also the boundaries of EWT, the boundaries of EWT and EFD are similar except the fourth boundary where EFD is closer to the third frequency spike than EWT. Moreover, 0 is a default boundary to FDM. The first segment of FDM is too narrow for the first component, and this deficiency induces the first two components to interfere with each other.

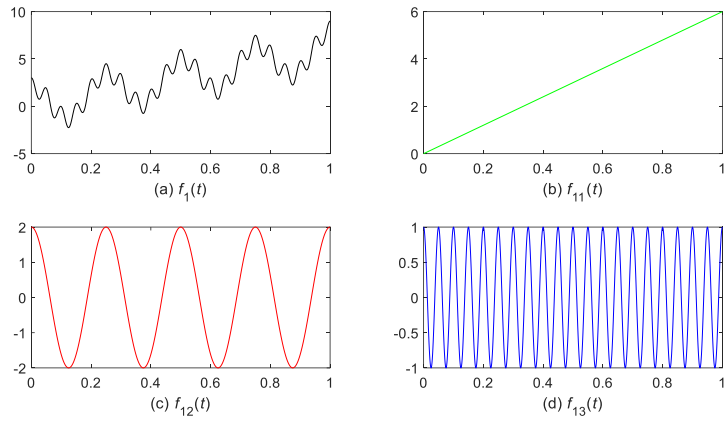

Fig. 4. $f_1(t)$ and its three components: from (a) to (d) is $f_1(t)$, $f_{11}(t)$, $f_{12}(t)$, and $f_{13}(t)$, respectively.

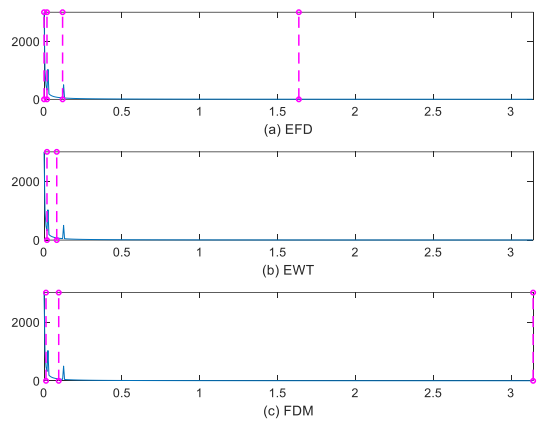

Fig. 5. Boundaries of $f_1(t)$ estimated by (a) EFD, (b) EWT, and (c) FDM.

Fig. 6 demonstrates the components extracted by EFD, EWT, and FDM, respectively. The results of EFD are shown in Fig. 6 (a), and the three modes reveal nice separation into a linear trend and two harmonics, respectively. However, segments of EWT are quite similar to EFD, as shown in Fig. 6 (b), the first component has a periodic fluctuate in mild because it is affected by the second component through the transition zone. Fortunately, the other two components are quite accurate extracted by EWT. Compared with the results of EFD and EWT, FDM shows no competitiveness in this example. As exhibited in Fig. 6 (c), except the third component, other results present weak performance. The second component has a mild linear trend, and the first just looks like a simple harmonic. Besides, distortions exist at both ends of the third. It is because FMD is unable to eliminate the boundary effect. For further purpose of comparison, the error is provided in Fig. 6 (d) which defined by the difference of an original component and the corresponding extracted component in time series. In this case, the errors of FDM are too large to compare, and only the errors of EFD and EWT are presented. From Fig. 6 (d), it is clear that the first component of the EFD has smaller errors than the EWT does. The other two components are in an opposite situation, but the errors of the EFD are close to the EWT.

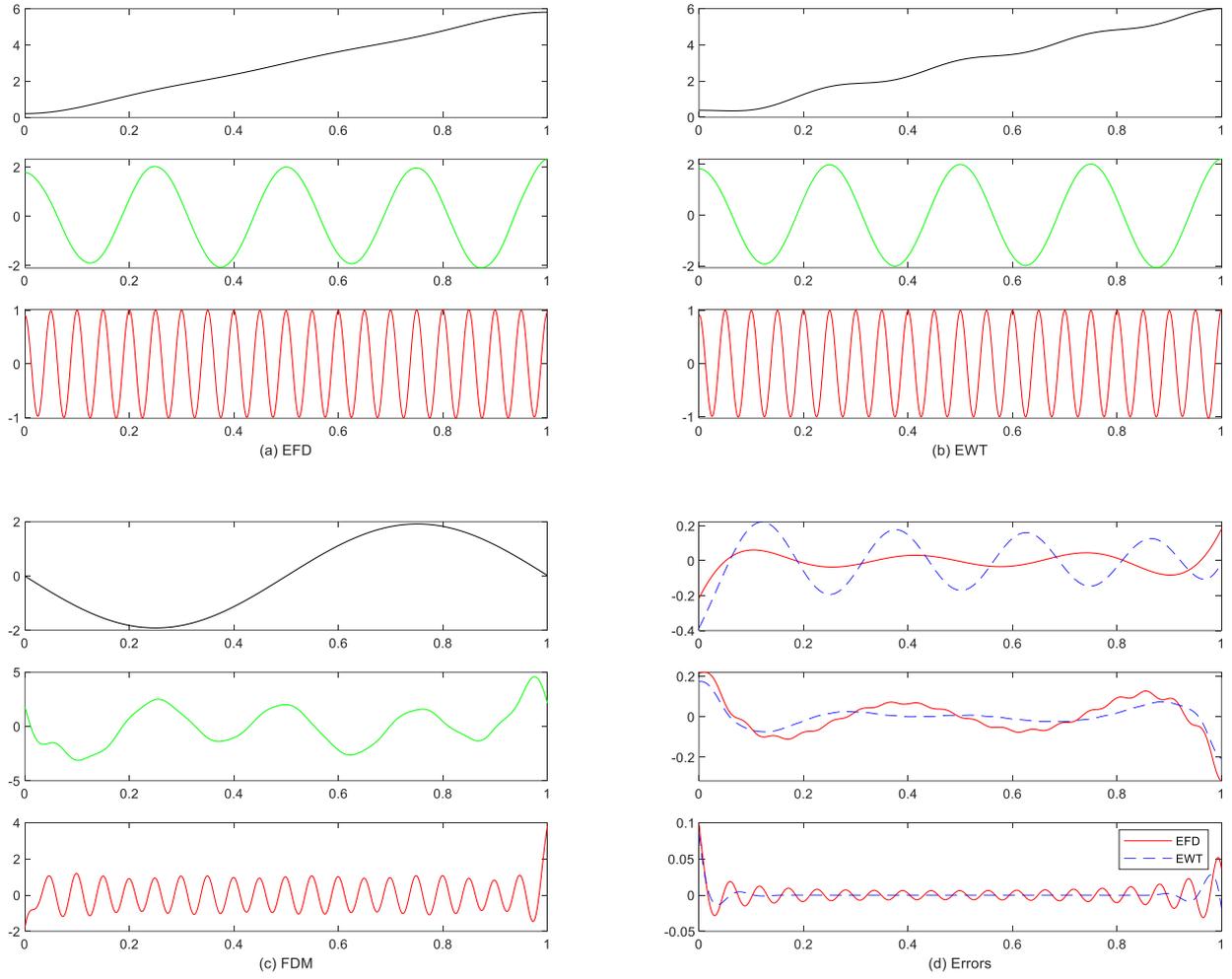

**Fig. 6. Decomposition results of $f_1(t)$ and their errors: (a) EFD, (b) EWT, (c) FDM, and (d) errors.**

b) Example 2: The second signal is a composition of a quadratic trend, a linear chirp signal, and a piecewise harmonic with two constant frequencies:

$$\begin{cases} f_{21}(t) = 6t^2 \\ f_{22}(t) = \cos(15\pi t + \pi t^2) \\ f_{23}(t) = \begin{cases} \cos(60\pi t) & \text{otherwise} \\ \cos(80\pi t - 15\pi) & \text{if } t \leq 0.5 \end{cases} \\ f_2(t) = f_{21}(t) + f_{22}(t) + f_{23}(t) \end{cases} \quad (23)$$

The time series of $f_2(t)$ and its components are demonstrated in Fig. 7. Then, decomposing the signal by exploiting the EFD, the EWT, and the FDM, respectively, and their boundaries are shown in Fig. 8. The number of segments of the FDM is three, but of the other two methods is four. This is because $f_{23}$ is departed by the EFD and the EWT, and it can be deemed as two independent components due to their frequencies have meaningful individual energy. However, the FDM keeps the entire $f_{23}$, and the first segment has the same limitation described in example 1.

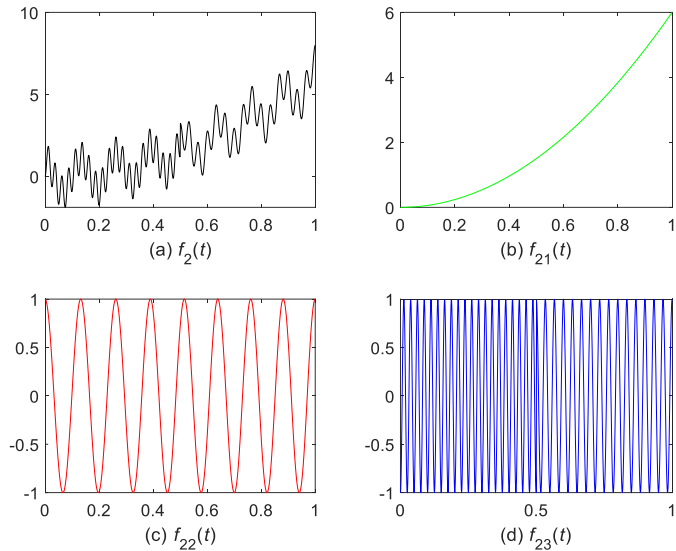

Fig. 7. $f_2(t)$ and its three components: from (a) to (d) is $f_2(t)$, $f_{21}(t)$, $f_{22}(t)$, and $f_{23}(t)$, respectively.

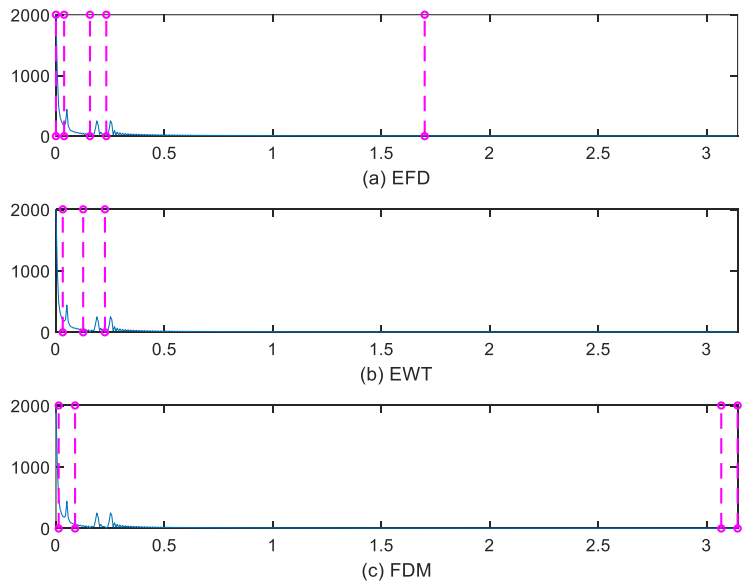

Fig. 8. Boundaries of $f_2(t)$ estimated by (a) EFD, (b) EWT, and (c) FDM.

Next, the components estimated via the EFD, the EWT and the FDM are displayed in Fig. 9. The results of the EFD are displayed in Fig. 9 (a), four components seem good. But the second component has some distortion in the middle, and the third and fourth components have not excellent performances in their transition zone. On the contrary, the better products are provided by the second component of the EWT as Fig.

9 (b) showed. However, some distractions occur in the front half portion of the third component, actually, the amplitude is zero in this part. Similar to example 1, it is clear depict in Fig. 9 (c) that the two first components of the FDM are not acceptable, and only the third component is satisfactory. In addition, Fig. 9 (d) shows the errors of the EFD and the EWT, it makes no difference in the first error. The second EFD's error is bigger in the middle, but the third error of the EFD is less than the EWT except for the middle transition zone.

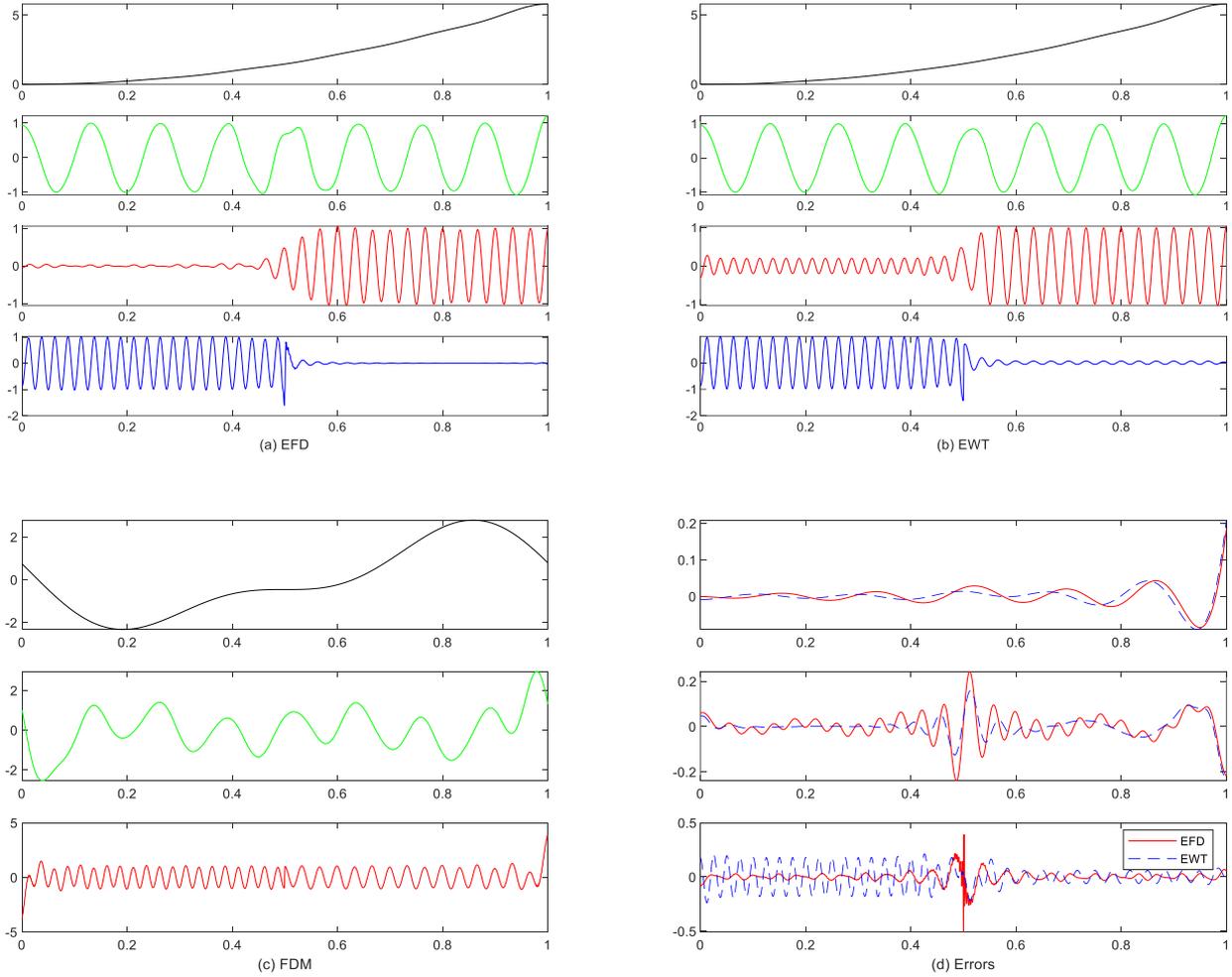

Fig. 9. Decomposition results of $f_2(t)$ and their errors: (a) EFD, (b) EWT, (c) FDM, and (d) errors.

c) Example 3: The third example embodies intra-wave frequency modulation:

$$\begin{cases} f_{31}(t) = \dfrac{1}{1.2+\cos(2\pi t)} \\ f_{32}(t) = \dfrac{\cos(32\pi t+0.2\cos(64\pi t))}{1.2+\sin(2\pi t)} \\ f_3(t) = f_{31}(t) + f_{32}(t) \end{cases} \quad (24)$$

Fig. 10 presents the time series of $f_3(t)$ and its components. Again, the EFD, the EWT, and the FDM are used to decompose $f_3(t)$, respectively. Their decomposing boundaries are shown in Fig. 11. In this case, the boundaries of the three approaches are basically identical except the last boundary of the EFD is closer to the third frequency spike. Hence, each component of the three methods is almost the same, as shown in Fig. 12. Nevertheless, the difference appears in the first component, the result of EWT and FDM has fluctuation after 0.6 s. Furthermore, because of the valuable results of FDM, the errors of these three methods are exhibited in Fig. 12 (d). The first error of the FDM seems large, but the errors are basically around 1.5. Because the entire amplitude-modulated appears in the first component. Therefore, the first component of FDM has the same form of $f_{31}(t)$. The second errors of these three methods are pretty small, among them, the errors of EFD are minimum.

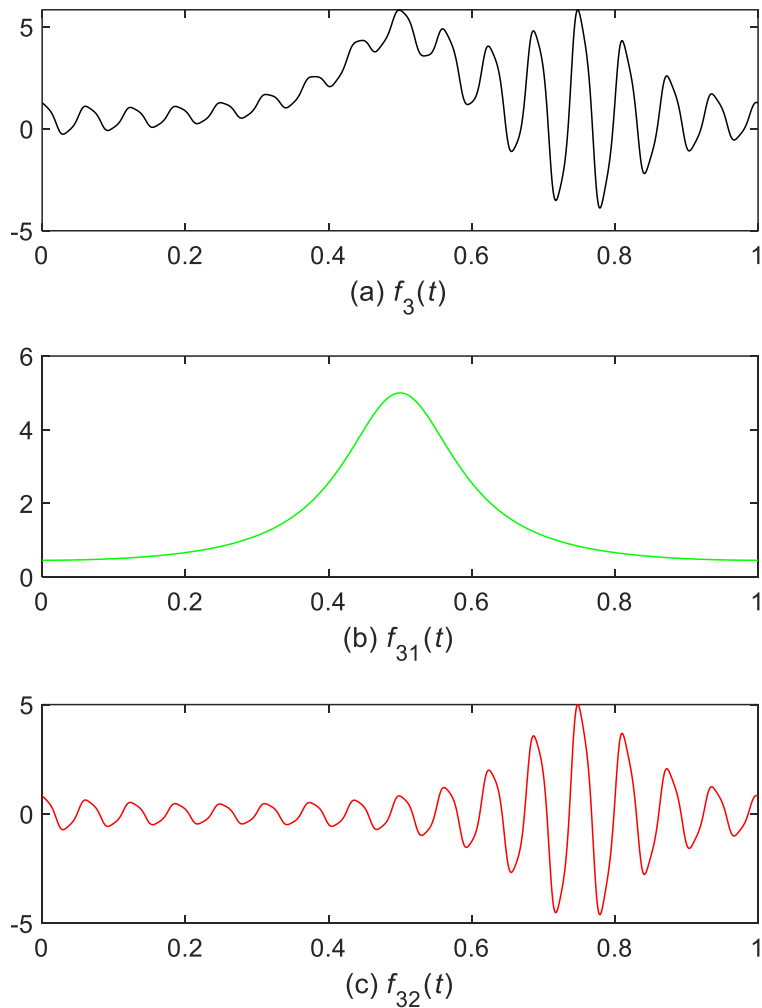

**Fig. 10.** $f_3(t)$ **and its three components: from (a) to (d) is** $f_3(t)$**,** $f_{31}(t)$**, and** $f_{32}(t)$**, respectively.**

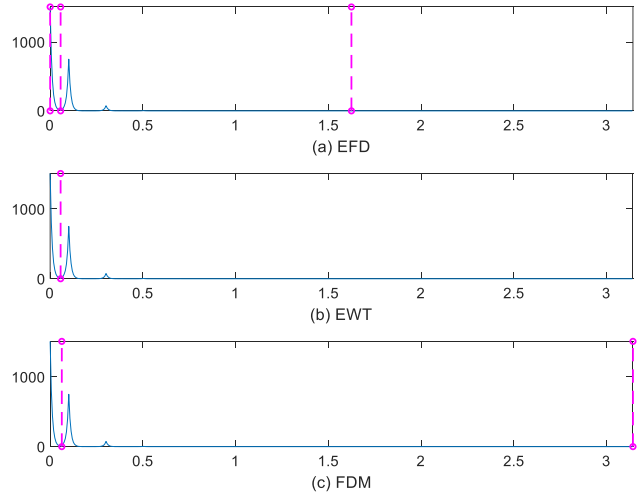

**Fig. 11.** Boundaries of $f_3(t)$ estimated by (a) EFD, (b) EWT, and (c) FDM.

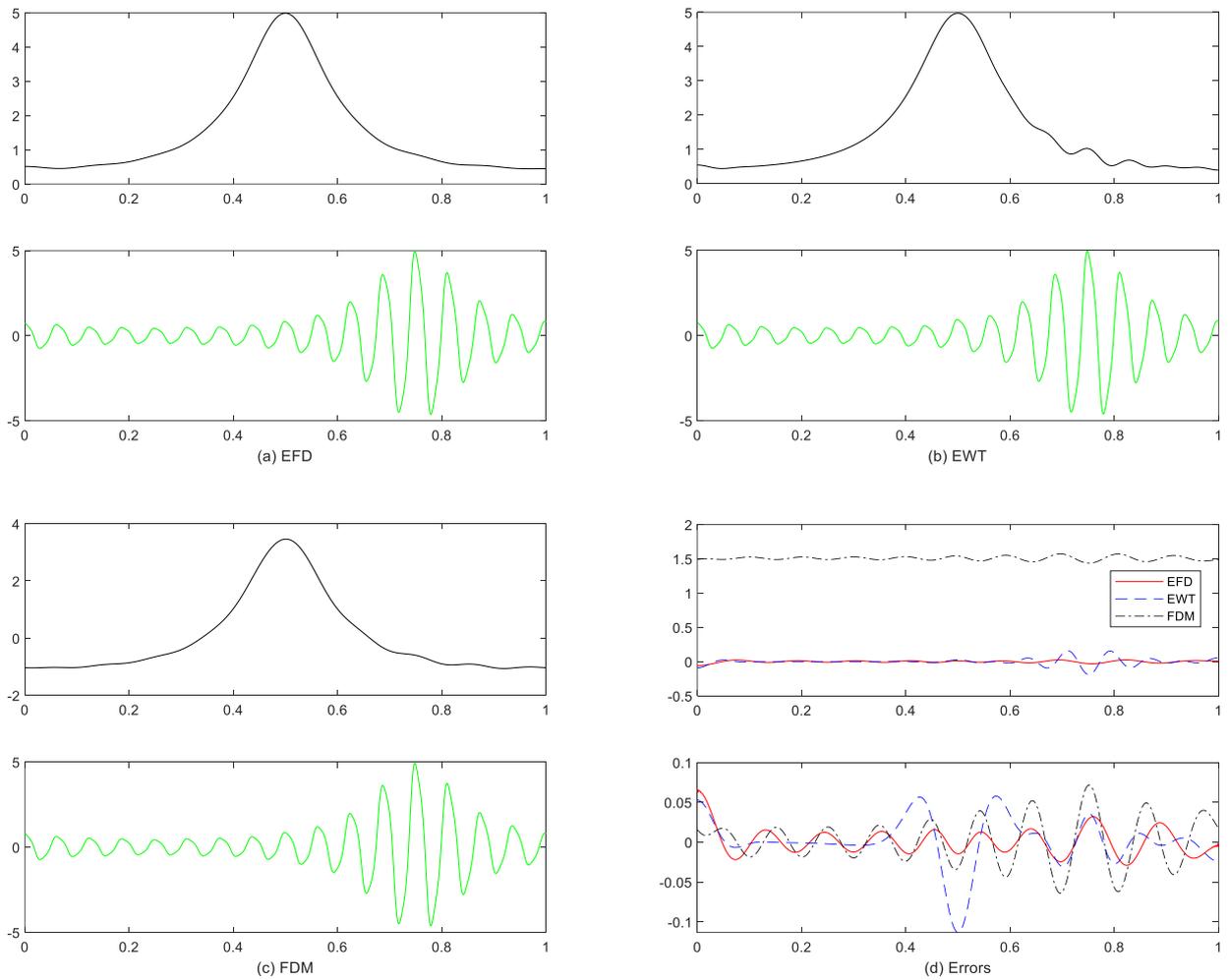

**Fig. 12.** Decomposition results of $f_3(t)$ and their errors: (a) EFD, (b) EWT, (c) FDM, and (d) errors.

*4.2 Simulated free vibration signal with two closely-spaced modes and a 20 dB noise*

To validate the capability of EFD in processing the closely-spaced modes and high-frequency noise, we apply a 3-degree of freedom structural free vibration response signal with varied damping ratios, two closely-spaced modes, and a 20 dB noise. It is utilized in [31] and given by the following equation:

$$s(t) = \sum_{i=1}^{3} A_i e^{-2\pi f_i \zeta_i t} \cos(2\pi t f_i \sqrt{1-\zeta_i^2} + \theta_i) + n(t) \tag{25}$$

where $A_i$ is the amplitude, $f_i$ is the natural frequency, $\zeta_i$ is the damping ratio and $\theta_i$ is the phase angle of the *i*th frequency, respectively. To simulate the free vibration response of a real structure, the low natural frequencies $f_1=1.1$, $f_2=1.3$, and $f_3=3.1$ Hz and the varied damping ratios $\zeta_1=2\%$, $\zeta_2=1.2\%$, and $\zeta_3=0.8\%$ are introduced. In addition, all of the amplitudes $A_i=1$, all of the phase angles $\theta_i=0$, and $n(t)$ is Gaussian noise with signal-to-noise-ratio of 20 dB. The synthetic signal sampled by a frequency of 50 Hz within a period of 20 s. Fig. 13 (a) and Fig. 13 (b) show the simulated signal with the noise of 0 dB and 20 dB, respectively.

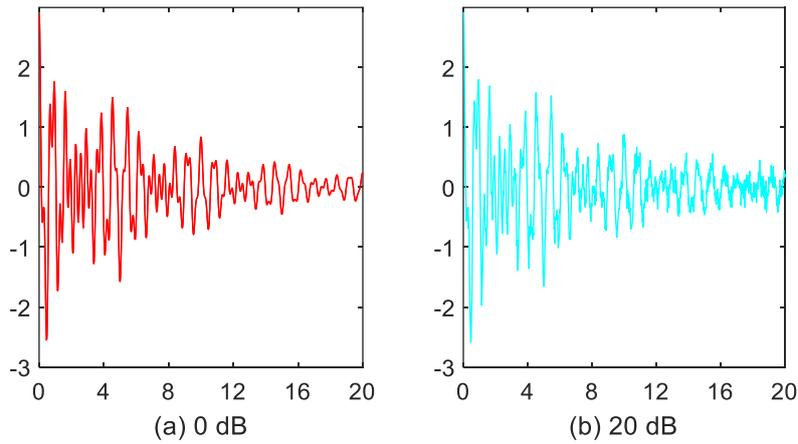

**Fig. 13. Synthetic signal: (a) 0 dB, (b) 20 dB.**

Results for extracting the boundaries are showed for three the proposal employing EFD, EWT, and FDM in Figs. 14 (a), 14 (b) and 14 (c), respectively. In this case, the segment technique of EWT is the local minima method [40], to eliminate the effect of differ boundary between EFD and EWT in closely-spaced modes. The number of segments in the methods of EFD, EWT, and FDM is 3, 4, and 13, respectively. However, the segments of EWT and FDM are not matched with the actual number 3. Each frequency spike contained in the first segment of EWT is not redundant, but the 10 extra segments of FDM are caused by Gaussian noise. The boundaries of EWT and EFD are same in the first three, but the fourth boundary of which EFD is closer to the third frequency spike than EWT. This difference means that the third component in EFD is less affected by high-frequency noise than EWT does. In addition, FDM is unable to separate the first two closely-spaced modes.

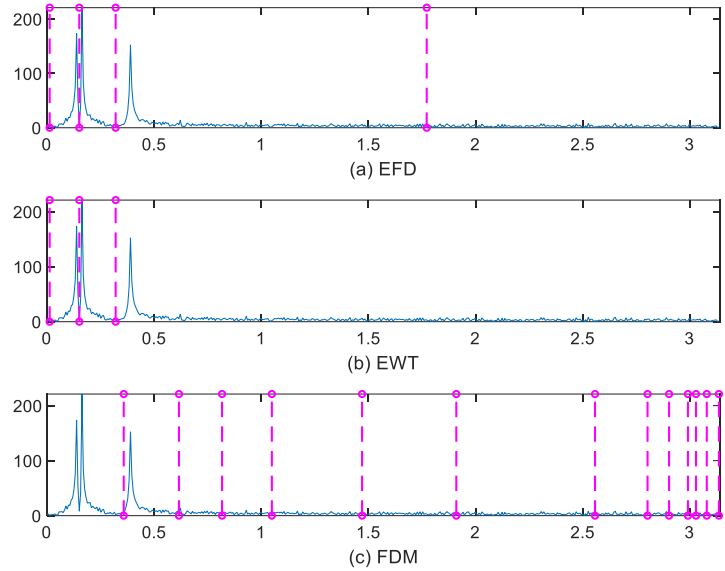

**Fig. 14. Boundaries of the simulated signal estimated by (a) EFD, (b) EWT, and (c) FDM.**

Next, the components of EFD and the 2nd, 3rd and 4th components of EWT and the 1st, 2nd components of FDM are presented in Fig. 15. The first two components of EFD have a high performance. Although the first two segments of EWT are same as EFD, the 1st and 2nd components interfere with each other. Meanwhile, the third embodies noise in both methods. The first component of FDM is disturbed by another frequency component, and the second component has no noise disturbance. Furthermore, the errors of EFD and EWT are exhibited in the box plots as Fig. 15 (d), which demonstrates that the errors of EFD are less than EWT.

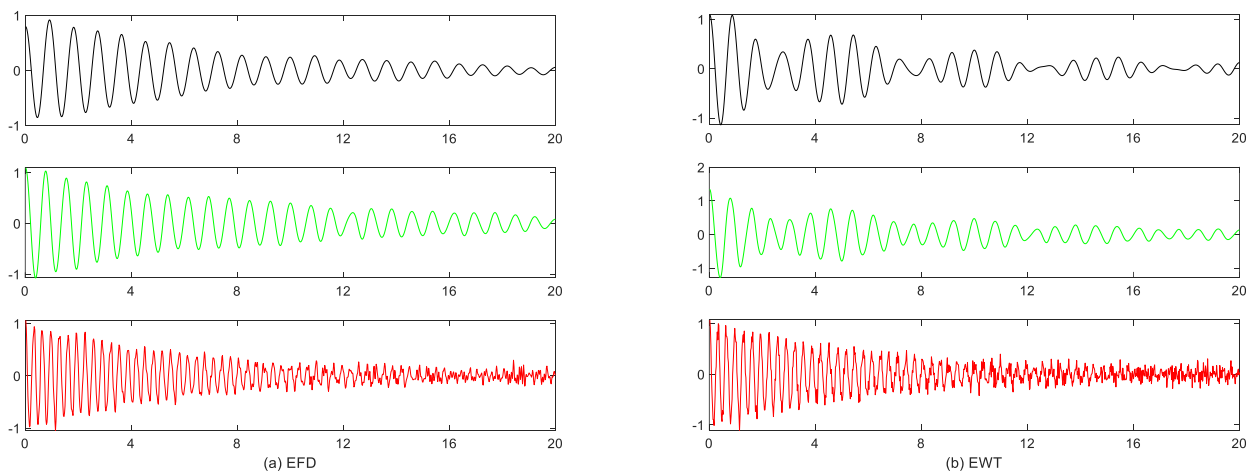

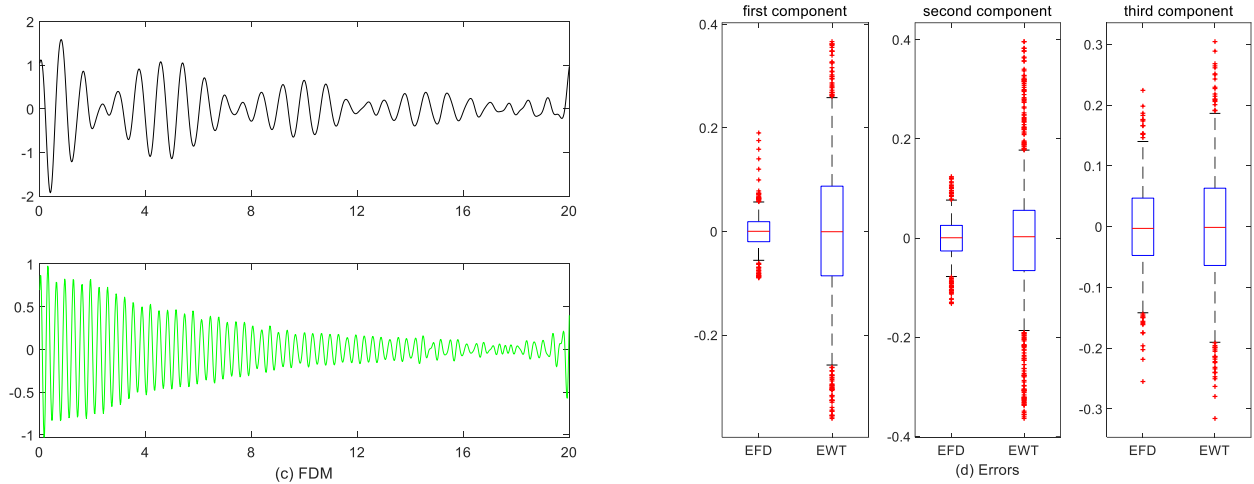

**Fig. 15.** Decomposition results of the synthetic signal and their errors: (a) EFD, (b) EWT, (c) FDM, and (d) errors.

*4.3 Real electrocardiogram signal*

To exploit the proposed method to real life, in this subsection, a real electrocardiogram (ECG) signal is introduced. This ECG signal is provided by the MIT-BIH arrhythmia database [43]. Data of 101 from 3600 to 4600 are used, and these 1000 samples are displayed in Fig. 16.

Because of substantial components, the results of EWT and FDM are just provided in the TF representation of the next subsection. Fig. 17 shows the boundaries extracted by the proposed method. There are 10 segments are obtained, and only one frequency spike contains in the first 9 segments. The decomposition results are presented in Fig. 18. The first component is the trend component, while the last one is the R wave. The other components are similar to the sinusoidal waves. More medical interpretation of these components should be remarked by cardiologists.

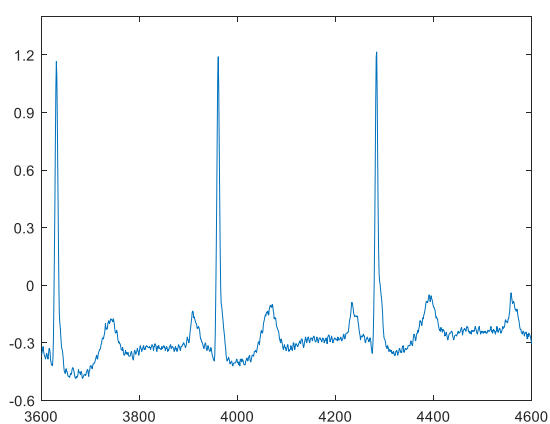

**Fig. 16.** A sample ECG signal.

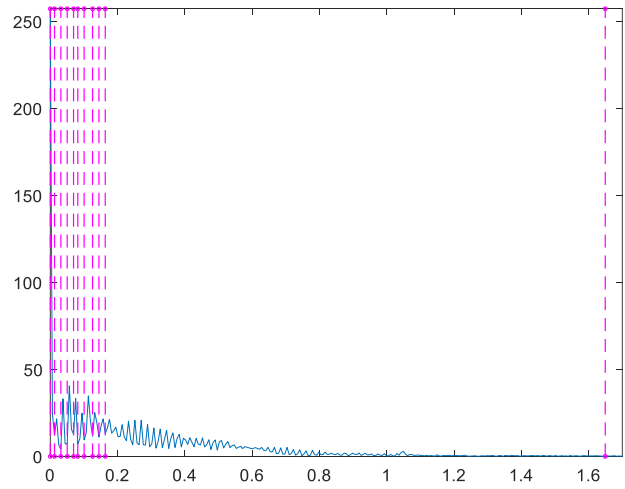

**Fig. 17. Boundaries of the ECG signal estimated by EFD.**

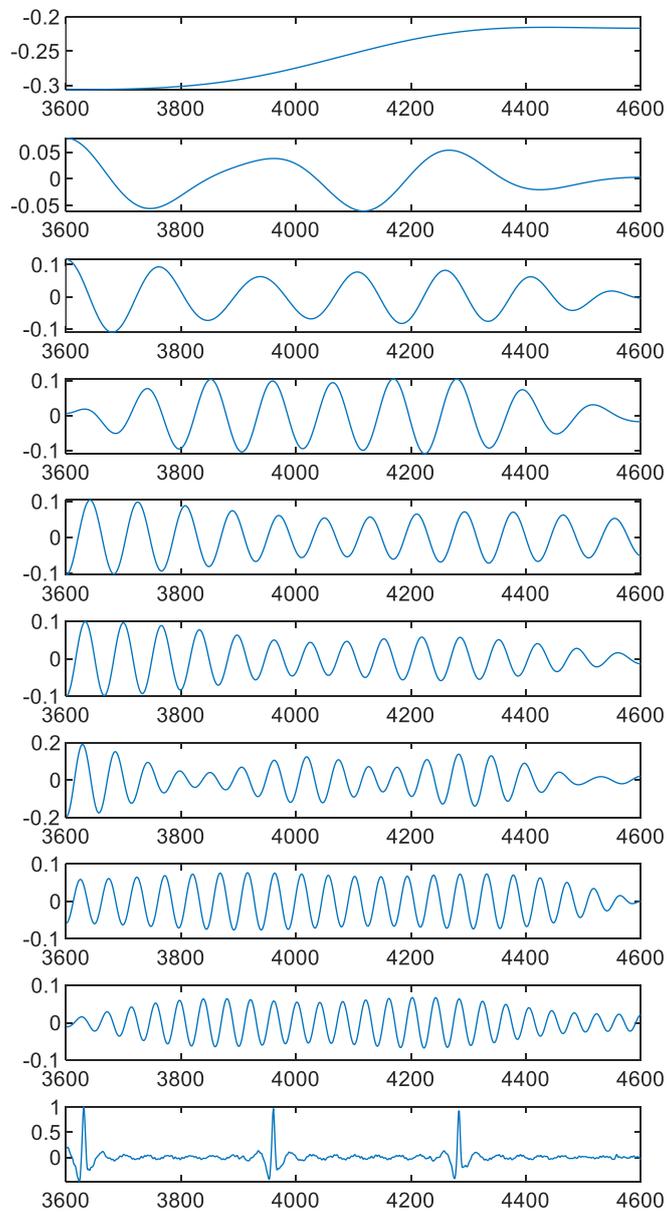

**Fig. 18. Decomposition results of the ECG signal by EFD.**

*4.4 Time-frequency representation*

Time-frequency representation is significant in signal processing, hence, the performance of EFD in the TF domain is exhibited in this section. In this case, $f_3(t)$ and the ECG signal are applied. EWT and FDM are also applied to these two signals for comparison.

Fig. 19 shows the TF representation of $f_3(t)$. The results of EFD, EWT, and FDM are demonstrated in Figs. 19 (a), 19 (b) and 19 (c), respectively. From these three figures, it manifests that each method contains two components and the corresponding component has the approximate same trend. For better comparison, Fig. 19 (d) presents the two components of $f_3(t)$ (the bottom is $f_{31}(t)$ and the top is $f_{32}(t)$) in the TF domain, respectively. Based on the results in Fig. 19 (d), EFD is more effective than the other two methods. On the one hand, the middle district of $f_{32}(t)$ of EWT fluctuates deeply. On the other, the energy of $f_{31}(t)$, as shown in Fig. 19 (d), is extremely tiny in the region of [0, 0.2] and [0.8, 1], but the results of EWT exists energy in the regions [0.8, 0.9] and the results of FDM exists energy in [0, 0.2] and [0.8, 1].

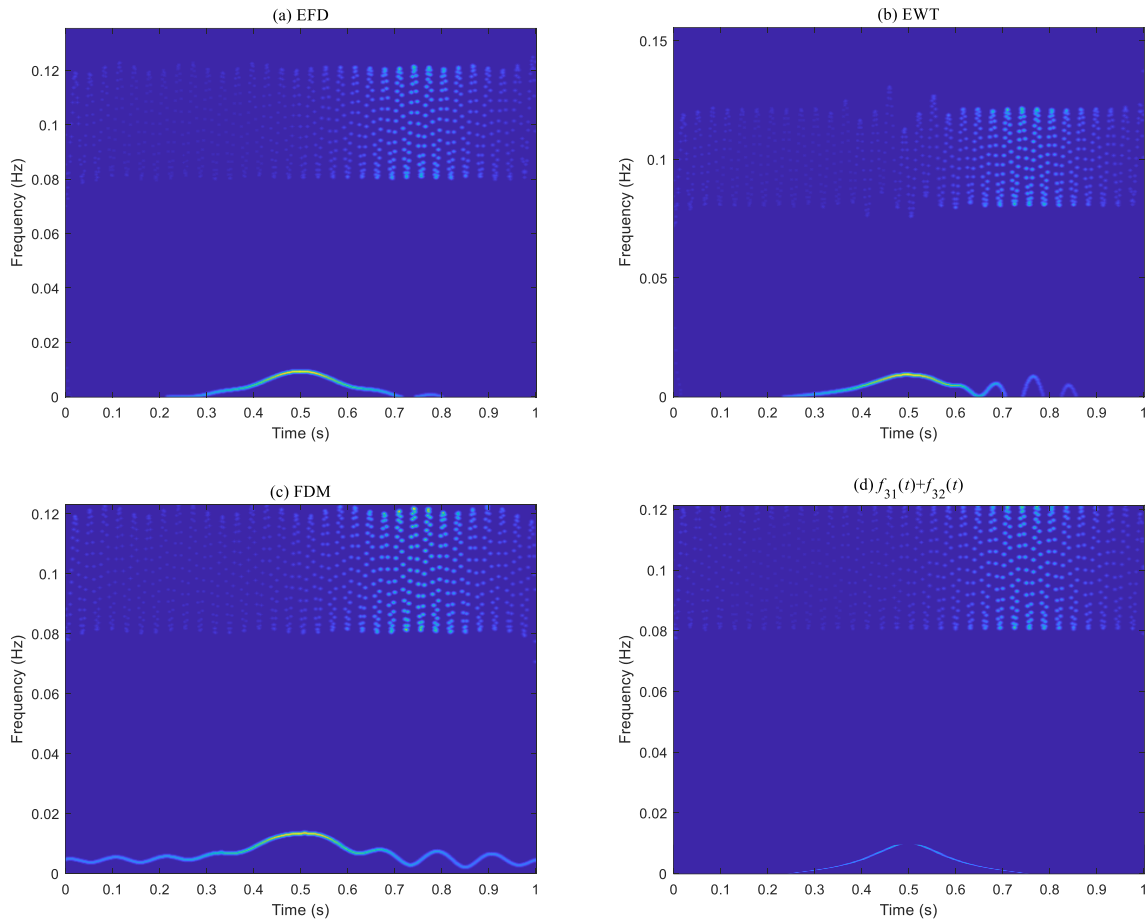

**Fig. 19. Time-frequency representation of $f_3(t)$ by: (a) EFD, (b) EWT, and (c) FDM; (d) the time-frequency representation of $f_{31}(t) + f_{32}(t)$ by the Hilbert transform.**

Then, TF representation of the ECG signal is shown in Fig. 20. Corresponding to the location of the R wave in the ECG signal, the energy of these methods are all concentrated in these sites. Moreover, the results of EFD and EWT are basically identical, but the energy of EWT between the three R waves is sparser than EFD. With regard to the results of FDM, it seems to provide a fruitful representation. However, it is worth noticing that FDM is interfered by noise in the region of relatively high-frequency.

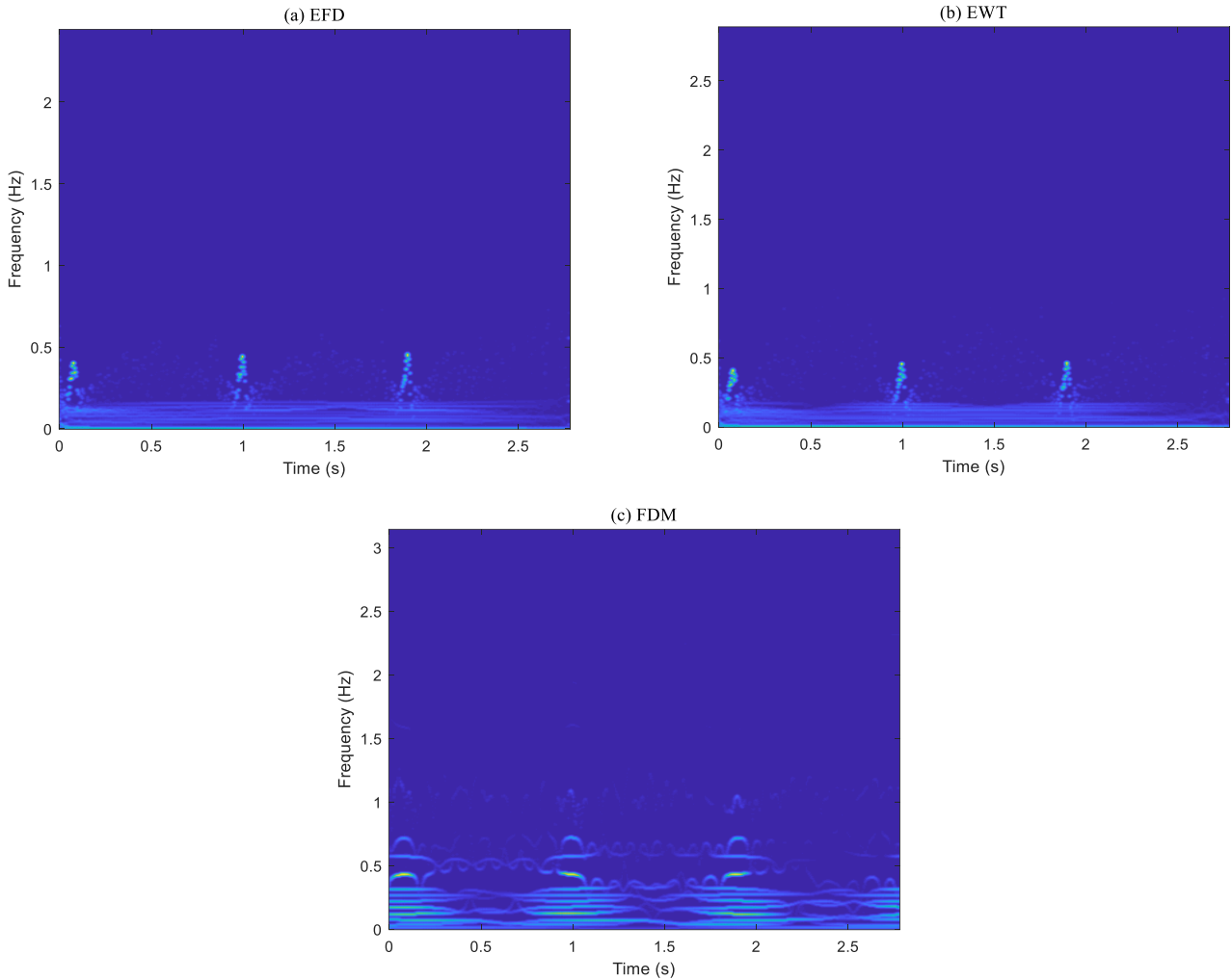

**Fig. 20. Time-frequency representation of the ECG signal by (a) EFD, (b) EWT, and (c) FDM.**

*4.5 Discussion*

Results showed in this paper indicate that EFD can provide good performance in the aforementioned examples. Firstly, the decomposition results of EFD are nearly the same as EWT for non-stationary multimode signals. But both of them present better than FDM. Secondly, EFD can decompose the closely-spaced frequencies and decrease the influence of noise for the last component. Finally, TF representation of EFD exhibits impressive performance.

The three methods: EFD, EWT, and FDM, all based on FFT. So, these methods have an equal time complexity $O(N\log_2 N)$ [44]. However, the proposed method EFD, unlike the other, has the characteristics of the simple framework and high computational efficiency.

Table 1 summarizes time consumption of these three methods extracting mono-component of aforementioned examples in this paper. All the calculations are performed on a laptop with an Intel Core i7-6700HQ CPU, 2.60GHz, 8.0GB RAM, Windows 10 Home, and software of MATLAB R2018a. Among the three methods, EFD requires less computational time. Furthermore, it is worth noticing that FDM consumes more computation time compared with the other. The time is increased, especially with the number of segments, as are shown in example 4 and 5.

**Table 1**

Computational time for the five examples.

| Example | Time (s) | | |
|---|---|---|---|
| | EFD | EWT | FDM |
| 1 | 0.0215 | 0.0409 | 0.8400 |
| 2 | 0.0259 | 0.0576 | 0.9237 |
| 3 | 0.0221 | 0.0435 | 0.7388 |
| 4 | 0.0260 | 0.0412 | 1.5323 |
| 5 | 0.0282 | 0.0497 | 2.0460 |

However, there are still restrictions of EFD which all leave room for further development:

1. The number of boundaries is required in advance and is one more than the signal model order.
2. Based on FFT, the proposed method is not appropriate for high-level noise signals.
3. The signal including components with overlapped in both the frequency and time domains is unable to be extracted.

## 5. Concluding remarks

In this paper, a novel approach is proposed based on EWT and FDM. The proposed method can be considered as a bandpass filter bank based on the Fourier spectrum segmentation. Meanwhile, the rigorous mathematical deduction based on FFT is also provided.

To substantiate the effectiveness of the proposed method, three non-stationary multimode signals, a closely-spaced frequencies signal, and an ECG signal are presented and tested. These experiments reveal that the proposed approach is capable of separate the aforementioned signals. Moreover, EWT and FDM are compared. The comparison revealed that EFD can separate the closely-spaced mode, while, typically, EWT exhibits that the extracted components cross each other. Another advantage of the proposed approach is that the last component of EFD is less influenced by high-frequency noise compared with EWT. On the other hand, comparing with FDM, the proposed method has higher processing precision, computation efficiency. Besides, the TF representation of $f_3(t)$ and ECG signal is presented, which indicate that EFD enables a better performance than the other does.

In the future, the proposed method would have wider applications in several fields. For instance, civil engineering, mechanical engineering, earthquake engineering, etc. Moreover, we would extend EFD to a higher dimension. For example, 2D-EFD will be able to process images. Meanwhile, the method of segmentation should be further developed, such as the power spectrum based method [15,16,31] and the adaptive method which likes machine learning technology [45]. It is noteworthy that each component of the signal is linearly superposed in the Fourier domain. If a signal contains the component with the energy spread over the whole Fourier spectrum or contains two consecutive modes are too close, these components would be interfered by the superposing effect. The reduction of the superposition effect should be regarded as a priority.

A MATLAB implementation of the proposed algorithm wil available at the MATLAB Central.

**Acknowledgments**

The authors would like to thank the people who share their contributions to the signal proceeding community.

**Conflicts of Interest:** The authors declare no conflict of interest.